\documentclass{article}

\usepackage{PRIMEarxiv}

\usepackage[utf8]{inputenc} 
\usepackage[T1]{fontenc}    
\usepackage{hyperref}       
\usepackage{url}            
\usepackage{booktabs}       
\usepackage{amsfonts}       
\usepackage{nicefrac}       
\usepackage{microtype}      
\usepackage{lipsum}
\usepackage{fancyhdr}       
\usepackage{graphicx}       
\graphicspath{{media/}}     
\usepackage{siunitx}
\usepackage{setspace}
\usepackage{amsmath}
\DeclareSIUnit\bp{bp}
\DeclareSIUnit\px{px}
\title{Tracking the Brownian motion of DNA-functionalized magnetic nanoparticles for conformation analysis beyond the optical resolution limit
}

\author{
  Christian Janzen, Yahya Shubbak, Rico Huhnstock,  Arno Ehresmann\\
  Institute of Physics and Center for Interdisciplinary Nanostructure Science and Technology (CINSaT) \\
  University of Kassel \\
  Heinrich-Plett-Straße 40, D-34132 Kassel, Germany\\
  \texttt{rico.huhnstock@physik.uni-kassel.de} \\
   \And
  Fabian Schmid-Michels, Inga Ennen, Andreas Hütten \\
  Faculty of Physics\\
  Bielefeld University\\
  Universitätsstraße 25, D-33615 Bielefeld, Germany\\
  \texttt{andreas.huetten@uni-bielefeld.de} \\
  \AND
  Melanie Wegener, Karl-Josef Dietz \\
  Faculty of Biology \\
  Bielefeld University\\
  Universitätsstraße 25, D-33615 Bielefeld, Germany
}

\begin{document}
\maketitle
\onehalfspacing

\begin{abstract}
Brownian motion provides access to hydrodynamic properties of nanoscale objects independent of their optical resolvability. Here, we present a diffusion-based approach to infer effective particle size distributions of DNA-functionalized magnetic nanoparticles (MNPs), consisting of a magnetic core and a polystyrene shell, in a regime where direct geometric sizing is limited by optical diffraction. Using multi-particle tracking microscopy, we analyze the Brownian dynamics of MNPs grafted with double-stranded DNA (dsDNA) of varying contour length under low-salt conditions.
A physically motivated model is introduced that relates dsDNA contour length to an effective hydrodynamic diameter via an attenuated corona description. The measured diffusion coefficient distributions exhibit a systematic and monotonic dependence on dsDNA length in quantitative agreement with the model. While the tracked objects are predominantly dsDNA-mediated agglomerates rather than isolated nanoparticles, clustering does not obscure the length-dependent signal. Instead, the dsDNA corona determines the hydrodynamic scaling, whereas agglomeration mainly introduces an offset and distribution broadening.
These results demonstrate that Brownian dynamics enables robust readout of biomolecular length scales even far below the optical resolution limit. The distribution-based approach is inherently tolerant to polydispersity and aggregation, making diffusion-based tracking a simple and promising strategy for future biotechnological and biomedical assays.
\end{abstract}


\section{Introduction}\label{sec:intro}

\begin{figure}[h!]
\centering
\includegraphics[scale = 0.9]{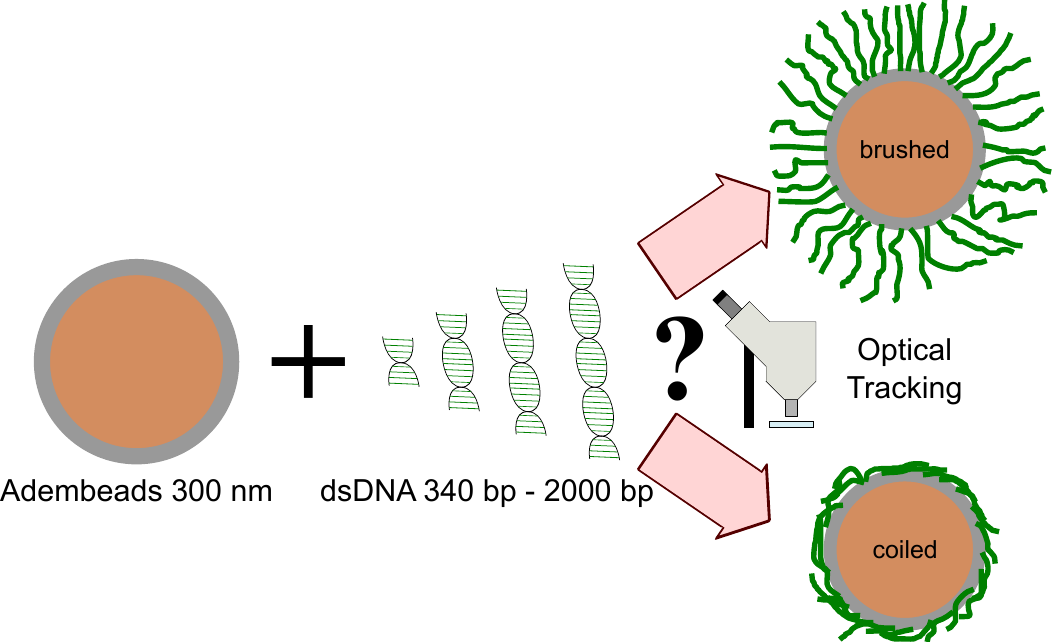}
\caption{Surface-functionalization of MNPs with dsDNA strands results either in a brushed or coiled conformation, depending on the DNA length and grafting density. Observation of the particle's Brownian motion via optical microscopy was used to distinguish between these states.}
\label{fig:1}
\end{figure}

DNA-functionalized colloidal particles in the nano- and micrometer size range are versatile tools for engineering self-assembled structures and play a pivotal role in detecting various biological and non-biological substances.~\cite{Rosi2005,Geerts2010,Chollakup2014} Pioneering studies successfully demonstrated the controlled bottom-up synthesis of larger structures from DNA oligo\-nucleotide-functionalized gold nanoparticles and nanocrystals.~\cite{Mirkin1996,Alivisatos1996} Detection schemes are reported for specific DNA-sequences~\cite{Maxwell2002}, metal ions~\cite{Liu2004}, enzymes~\cite{Patolsky2003}, and other analyte species~\cite{Liu2006} using nanoparticles or quantum dots hybridized with short DNA strands as probes. A steady interest can be noted in integrating magnetic nano- and microparticles for separation, fluid mixing, drug delivery, and hyperthermia by magnetic field-induced particle motion.~\cite{Pankhurst2003,VanReenen2014,Rampini2021, Holzinger2012, Holzinger2015, Elschner2024,Huhnstock2026} This technique also allows novel methods for analyte binding detection, like surface interaction-mediated traveling wave magnetophoresis~\cite{Shubbak2026}, magnetic particle spectroscopy~\cite{Biederer2009}, and magnetic particle aggregation assays~\cite{Ran2014}, aiding the development of point-of-care medical diagnostics.\\
For DNA-functionalized magnetic particles, the concentration of attached DNA, also known as the grafting density, is of major importance for their physical properties. For high grafting densities, DNA strands typically form a rigid polymer brush around the particle, enabling binding of a specific target in every direction.~\cite{Geerts2010} The DNA brush at high grafting densities significantly changes the particle's hydrodynamic radius and, therefore, impacts its dynamic response to an applied magnetic field when suspended in a liquid.~\cite{Lak2024} At low grafting densities, the conformation of particle-bound DNA strands is coiled, where single strands are wrapping around the particle. This effectively increases the probability of non-specific binding of DNA on the particle's surface as compared to the brushed state.~\cite{Geerts2010,Lak2024} A mixture of coiled and brushed single-stranded DNA (ssDNA) attached to magnetic nanoparticles was found to yield the highest sensitivity in magnetic particle spectroscopy-based biosensing of target DNA sequences.~\cite{Lak2024}\\
Experimental validation of the conformation of DNA bound to magnetic nano- or microparticle surfaces is, thus, essential for understanding their binding affinity toward specific targets. Atomic force microscopy~\cite{Binnig1986} is a possible imaging technique, providing high spatial resolution, but typically yields a low throughput and is hard to apply in situ. The low throughput issue might be solved by using either magnetic particle spectroscopy~\cite{Biederer2009,Rsch2021,Lak2024} or dynamic light scattering (DLS).~\cite{Stetefeld2016} The former, however, is only applicable for magnetic DNA-functionalized particles and solely probes the dynamic response of the particles to an alternating current (AC) magnetic field, which does not necessarily reflect the quasi static, undisturbed state of DNA conformation at the particle's surface. DLS is indeed useful to investigate the hydrodynamic radius of particles, which in turn gives information about the structure of particle-bound DNA, but is prone to overestimating the weight of larger particles or aggregates in the obtained size distribution, is highly dependent on the carrier liquid properties, and is in general a low-resolution technique.~\cite{Stetefeld2016} Other methods to determine the hydrodynamic radius of DNA-functionalized particles include the capture inside an optical trap with low stiffness and subsequent analysis of the particle's Brownian motion.~\cite{Ueberschr2011} Here as well, the low measurement throughput is not ideal, investigating only a single particle at a time. Finally, the free end of a particle-bound DNA strand can be attached to a container wall, and subsequent observation of the particle's restricted Brownian motion can be used to monitor changes in the conformation of the attached DNA strand. This is also known as tethered particle motion.~\cite{Brinkers2009,Fan2018} While delivering accurate results for the behavior of single strands, it does not cover the combined structure of many DNA strands immobilized on the particle's surface. It also struggles with low throughput, as only a single particle is observed per experiment. \\
This work presents an efficient yet simple experimental approach for characterizing the structural conformation of short, double-stranded (ds) DNA bound to the surface of spherical magnetic nanoparticles (MNPs). The basic idea behind our study is sketched in Figure~\ref{fig:1}. By combining optical brightfield microscopy with a customized multiple particle tracking algorithm to observe the free two-dimensional Brownian motion of MNPs, we obtained data on the particles' diffusivity with high statistical significance. The resulting diffusion coefficient distributions are then connected to effective hydrodynamic diameters through the Stokes–Einstein relation. From the measured distribution of diffusivity, the effective hydrodynamic diameter distribution is reconstructed inversely. In parallel, we analyze the bare MNPs by transmission electron microscopy (TEM) and find that the nominally monodisperse particles actually follow a lognormal size distribution. We use this experimentally determined distribution as the starting point for a forward hydrodynamic model based on an attenuated hedgehog representation of the dsDNA corona at low ionic strength. Both the coiled state and the brushed state (Figure~\ref{fig:1} right) are generally representable by the attenuated hedgehog model. Hence, we can conclude the structural conformation of dsDNA with varying length (between 340 base pairs (bp) and \SI{2000}{\bp}). By combining the forward modeling from TEM-based size distributions and the inverse reconstruction from the measured diffusion coefficients, we obtain a closed picture. On the one hand, the attenuated hedgehog model predicts how effective diameter and diffusion coefficient should evolve with dsDNA length. On the other hand, the optical tracking experiments tell us what actually happens in dispersion. The key result is that both routes lead to the same dependence of hydrodynamic size on dsDNA length, even though the tracked objects turn out to be dsDNA-mediated particle agglomerates rather than individual dsDNA-bound MNPs far below the optical resolution limit. As we analyze the diffusivity of many MNPs (more than 100) within the same batch, we account for the fabrication-related size variation of the particles, which limits other methods, e.g., DLS. This finding potentially enables a high-throughput in situ structural characterization technique for particle-bound substances (not only DNA) based on a common experimental setup. For instance, live detection of folding events for particle-bound proteins might be within reach, with no restriction to measurement sensitivity related to the intrinsic particle size distribution.

\section{Theory}\label{sec:theory}
\subsection{Brownian motion, mean squared displacement and the Stokes-Einstein relation}\label{subsec: bm, msd, ses}

Assuming a perfect random walk-like motion of the MNPs in the overdamped, low-Reynolds-number regime, their Brownian motion projected into the imaging plane $\left(xy\right)$ can be quantified using the mean squared displacement (MSD) over lag-time $\tau$~\cite{Catipovic2013}.
\begin{equation}
\mathrm{MSD}\left(\tau\right) = \langle \Delta x^2\rangle\left( \tau\right) + \langle \Delta y^2\rangle\left( \tau\right) = \langle \Delta r^2\rangle\left( \tau\right) = 4D\tau,
\label{eq:1}
\end{equation}

with the diffusivity $D$ of an object actuated by thermal fluctuations of the surrounding fluid.~\cite{Einstein1905, Chandrasekhar1943} It is a function of the fluid temperature $T$, the fluid viscosity $\eta$, and the hydrodynamic particle radius or diameter $r_{\mathrm{h}}$, $d_{\mathrm{h}}$ (assuming perfect spherical shape of the particle)~\cite{Einstein1905, Catipovic2013}:
\begin{equation}
D=\frac{k_{\mathrm{B}}T}{6\pi\eta r_{\mathrm{h}}} =\frac{k_{\mathrm{B}}T}{3\pi\eta d_{\mathrm{h}}}=\frac{C}{d_{\mathrm{h}}},~\mathrm{with}~ C=\frac{k_B T{}}{3\pi\eta}.
\label{eq:2}
\end{equation}
For $T = \SI{298.15}{K}$ and $\eta = \SI{0.89}{mPa~s}$ one finds $C\approx\SI{491e{-21}}{ m^3~s^{-1}}$.

\subsection{Lognormal diffusion and diameter distributions}\label{subsec: lognorm. diff, diam}
TEM analysis of bare (no dsDNA) streptavidin-coated MNPs streptavidin-coated MNPs (Ademtech 0313, nominal diameter 300 nm) shows that the core diameters follow a lognormal distribution. Likewise, the experimentally determined
diffusion coefficient distributions of bare and dsDNA-functionalized MNPs are well described by lognormal functions.
We therefore represent the diffusion coefficient distribution $f_D\left(D\right)$ as

\begin{equation}
\label{eq:f_D(D)}
    f_D\left(D\right)= \frac{1}{D\sigma_D\sqrt{2\pi}}\exp\left[-\frac{\left(\ln{D}-\mu_D\right)^2}{2\sigma_D^2}\right],
\end{equation}
where $\mu_D$ and $\sigma_D$ denote the mean and standard deviation of $\ln{D}$, respectively.
Because $D$ and the effective diameter $d$ are deterministically related via Equation~\ref{eq:f_D(D)}, the probability density of effective diameters $f_d\left(d\right)$ can be obtained from $f_D\left(D\right)$ by a change of variables. Using $D\left(d\right) =C/d$ and $\left|\mathrm{d}D/\mathrm{d}d\right| =C/d^2$, we obtain with Equation~\ref{eq:f_D(D)} and $\mu_d = \ln{C}-\mu_D$ and $\sigma_d = \sigma_D$ a lognormal distribution for $d$:

\begin{equation}
\label{eq:f_d(d)}
    f_d\left(d\right)= \frac{1}{d\sigma_D\sqrt{2\pi}}\exp\left[-\frac{\left(\ln{d}-\left(\ln{C}-\mu_D\right)\right)^2}{2\sigma_D^2}\right].
\end{equation}

In what follows, we use this relationship in two directions: to predict diffusion coefficient distributions from a known diameter distribution, and to reconstruct effective diameter distributions from measured diffusion coefficient distributions.

\subsection{Attenuated hedgehog model for dsDNA coronas}\label{subsec: att. hedgehoc}
In the experiments described below, the dsDNA-functionalized MNPs are suspended in deionized water at very low ionic strength ($I \leq  10^{-4}$\textsc{m}). Under these conditions, the Debye length is on the order of several tens of nanometers, comparable to or larger than the DNA double-helix radius. Electrostatic repulsion along each molecule and between neighboring strands favors conformations in which the dsDNA extends away from the MNPs surface. This naturally suggests a hedgehog-like picture of the corona.

The contour length $L_{\mathrm{c}}$ of dsDNA with $N_{\mathrm{bp}}$ base pairs is

\begin{equation}
    \label{eq:Lc}
    L_{\mathrm{c}}=aN_{\mathrm{bp}}, \quad a =0.34~\mathrm{nm}~\mathrm{bp^{-1}},
\end{equation}
where $a$ is the rise per base pair for B-form DNA. In an idealized “full hedgehog” model one would expect
\begin{equation}
    \label{eq:full_hedgehog}
    d_{\mathrm{eff}}^{\left(\mathrm{full}\right)}=d_0+2 L_{\mathrm{c}},
\end{equation}

with $d_0$ being the diameter of the bare particle and $d_{\mathrm{eff}}^{\left(\mathrm{full}\right)}$ the diameter of the particle fully covered by DNA, assuming that the DNA is not bent. In reality, thermal bending, incomplete radial alignment, and local steric constraints reduce the mean radial extension. To capture this effect, we introduce an attenuated hedgehog model
\begin{equation}
    \label{eq:attenuated_hedgehog}
    d_{\mathrm{eff}}^{\left(\mathrm{att}\right)}=d_0+2\gamma L_{\mathrm{c}},
\end{equation}

with $0<\gamma<1$. Guided by earlier studies on surface-tethered dsDNA and polyelectrolyte brushes,\cite{Rivetti1996,RomeroSanchez2022,Besteman2004,Ballauff2006} we use $\gamma \approx 0.6$ as a realistic estimate for the average radial projection under low-salt conditions. Starting from the TEM-derived lognormal diameter distribution of bare MNPs $f_{d_0}\left(d\right)$, the effective diameter distribution for a given dsDNA length is then obtained by a simple shift,
\begin{equation}
    \label{eq:f_d_eff}
    f_{d_\mathrm{eff}}\left(d\right)=f_{d_0}\left(d-2\gamma L_{\mathrm{c}}\right),
\end{equation}
with $f_{d_\mathrm{eff}}\left(d\right)=0$ if $d-2\gamma L_{\mathrm{c}}\leq 0$. From these distributions, we compute model diffusion coefficient distributions via Equation~\ref{eq:2}.

\subsection{Optical resolution}\label{subsec: opt. res.}
The optical tracking experiments are performed with a numerical aperture $NA = 0.8$ and white-light LED illumination in the wavelength range $\lambda = 420-700~\mathrm{nm}$. The lateral Abbe diffraction limit is given by
\begin{equation}
    \label{eq:Abbe}
    \delta_{xy}= \frac{0.61 \lambda}{NA}\approx320-530~\mathrm{nm},
\end{equation}

depending on the wavelength.~\cite{Israelachvili2011} Bare MNPs with effective diameters of the order of \SI{165}{nm} are therefore well below the resolution limit and appear as point-spread-function (PSF) limited spots. Geometric size differences in this regime cannot be resolved directly; the relevant size information resides in the Brownian dynamics.\\
In the model plots of effective diameter distributions, we indicate the unresolved region as a black and gray band. This makes it clear at a glance which parts of the distributions fall below (black), inside (gray), or above (white) the optical resolution window.

\section{Results}\label{sec:results}
\begin{figure}[h!]
\centering
\includegraphics[scale = 0.9]{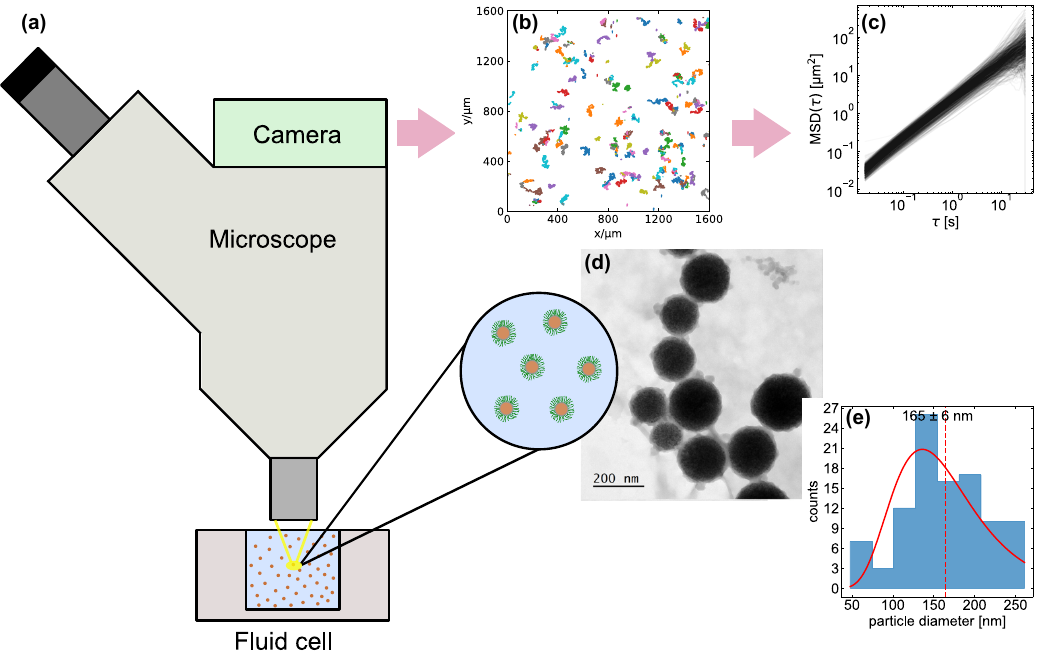}
\caption{Sketch of the experiment and data evaluation. (a) An optical bright-field microscope equipped with a high-resolution camera was used to record the motion of dsDNA-functionalized MNPs that were free-floating in a container cell. (b) A particle tracking algorithm provides 2D trajectories for each MNP visible in the field of view (FOV). (c) The mean squared displacement (MSD) was calculated for each particle trajectory and plotted as a function of lag time $\tau$, denoting different time frames for trajectory analysis. (d) The obtained MSD is closely connected to the size distribution of observed MNPs, which was examined via transmission electron microscope (TEM) images. (e) Histogram for the measured MNP diameters obtained from TEM images, fitted by a log-normal distribution function. A mean particle diameter of $165\pm6~\mathrm{nm}$ was obtained.}
\label{fig:2}
\end{figure}

Strands of dsDNA with varying length (\SI{340}{\bp}, \SI{500}{\bp}, \SI{722}{\bp}, and \SI{2000}{\bp}) were grafted onto commercially available Bio-Adembeads Streptavidin with a nominal diameter of \SI{300}{\nano\meter}. Using the distance per bp for B-form DNA (\SI{0.34}{\nano\meter})~\cite{Seol2007}, the contour (stretched-out) lengths of the employed DNA fragments translate to a range of \SI{116}{\nano\meter} to \SI{680}{\nano\meter}. Refer to the Methods section for details on the DNA strand synthesis and grafting procedure. The DNA-functionalized MNPs were suspended in deionized water and transferred to the well of a standard microtiter plate. Each well containing a particle type (given by the respective DNA length) was placed under an optical bright-field microscope (see experimental setup in Figure~\ref{fig:2}~(a)).

\subsection{TEM size distribution of bare MNPs and forward modeling}\label{subsec: TEM naked model}
A portion of the investigated MNPs was analyzed with transmission electron microscopy (TEM) to experimentally determine the size distribution. An exemplary TEM image of several bare MNPs (no DNA functionalization) is shown in Figure~\ref{fig:2}~(d). From these measurements, we inferred a mean MNP diameter of $165\pm6~\mathrm{nm}$. This is significantly smaller than the nominal diameter of \SI{300}{\nano\meter} given by the distributor. A histogram of the measured MNP diameters is given in Figure~\ref{fig:2}~(e). The red solid line represents a fit of the histogram with a log-normal distribution function. The MNPs are clearly polydisperse, with a mean diameter of $d_{0\mathrm{,mean}}=165\pm6~\mathrm{nm}$ (red dashed line) and a pronounced right-skewed tail. This lognormal diameter distribution $f_{d_0}\left(d\right)$ serves as the starting point for the forward model.\\

\begin{figure}[h!]
\centering
\includegraphics{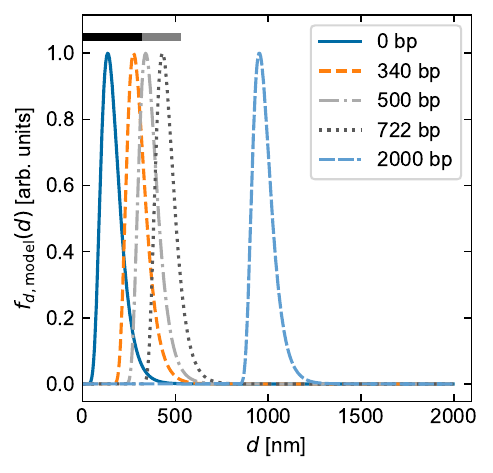}
\caption{Predicted effective diameter distributions $f_{d,\mathrm{model}}\left(d\right)$ for dsDNA-functionalized MNPs using the attenuated hedgehog model with $\gamma = 0.6$. The gray band denotes the Abbe-diffraction-limited regime for the given numerical aperture and wavelength range. Within the black band, (dsDNA)-MNPs fall below the Abbe limit.}
\label{fig:d_model}
\end{figure}

Applying the attenuated hedgehog model with $\gamma = 0.6$ to $f_{d_0}(d)$ yields, for each dsDNA length, the predicted effective diameter distributions $f_{d_{\mathrm{model}}}\left(d\right)$ shown in Figure~\ref{fig:d_model}. The peaks of the distributions shift steadily to larger diameters as the dsDNA length increases. The Abbe-limited resolution window is indicated as a black and gray band. Bare MNPs and MNPs with shorter dsDNA predominantly fall below or within this window and appear point-spread-function limited; only for the longer dsDNA (\SI{720}{\bp} and \SI{2000}{\bp}) does a significant fraction of the distribution move into the optically resolvable regime.

\begin{figure}[h!]
\centering
\includegraphics{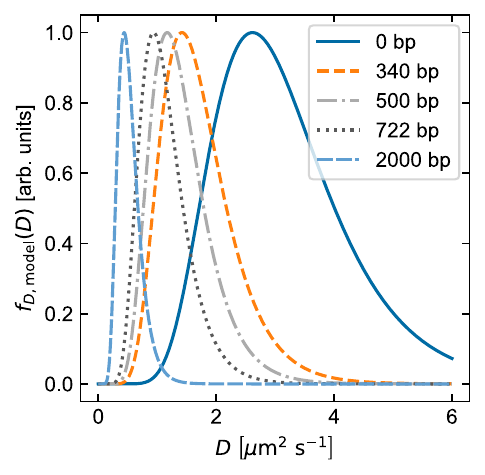}
\caption{Model diffusion coefficient distributions $f_{D,\mathrm{model}}\left(D\right)$ derived from the effective diameter distributions in Figure~\ref{fig:d_model} via the Stokes–Einstein relation. Longer dsDNA systematically shifts the distributions to lower D.}
\label{fig:D_model}
\end{figure}

Using Equation~\ref{eq:2}, we convert these effective diameter distributions into model diffusion coefficient distributions $f_{D,\mathrm{model}}\left(D\right)$ (Figure~\ref{fig:D_model}). As expected, the peaks move to lower $D$ as the dsDNA becomes longer.

\subsection{Measured diffusion coefficient distributions}\label{subsec: meas diff coeff}
\begin{figure}[h!]
\centering
\includegraphics{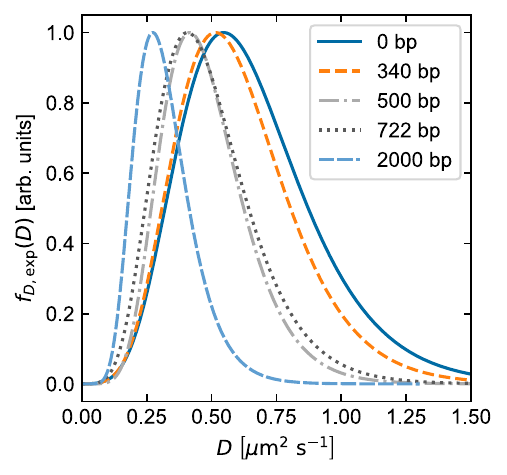}
\caption{Experimental diffusion coefficient distributions $f_{D,\mathrm{exp}}(D)$ obtained from optical tracking for bare MNPs and MNPs functionalized with 340, 500, 720, and \SI{2000}{\bp} dsDNA. Lognormal fits
(lines) provide the basis for inverse reconstruction of effective diameters shown in Figure~\ref{fig:d_exp}.} 
\label{fig:D_exp}
\end{figure}

The free Brownian motion of the MNPs in the surrounding static liquid was recorded using a high-resolution camera (DMK 33UX265 with Sony IMX265 sensor) for a duration of \SI{120}{\sec} at 60 frames per second (fps). The field of view (FOV) in the microscope ($\SI{2048}{\px}\,\times \,\SI{1536}{\px}$, translating to $\SI{355.06}{\micro\m} \,\times \,\SI{266.30}{\micro\m}$) was chosen to be far away from the container walls of the well; therefore, observing a non-restricted motion of MNPs in 3D. Due to sedimentation and Brownian motion, some MNPs were entering and leaving the FOV during the experiment. It was ensured that only those MNPs were analyzed that were consistently present in the FOV. Consequently, the Brownian motion can be treated as two-dimensionally free.\\
Next, the trajectory of each particle visible in a single video was analyzed using the Python-based code library Trackpy 0.7~\cite{Trackpy}. It enables the automatic tracking of multiple particles visible as dark/bright spots against a contrasting background (more details in the Methods section). Here, we focus on the trajectory analysis of each MNP in the 2D plane parallel to the microscope's focal plane. In general, the MNPs perform Brownian motion in all three dimensions; however, we can only access the two-dimensional in-plane motion in our current analysis. An all-optical 3D tracking of particles is methodologically possible~\cite{Barnkob2021, Huhnstock2022} and may be applicable in future studies. A typical tracking result, obtained for MNPs without DNA attachment, is shown in Figure~\ref{fig:2}~(b). The tracking results in $x$ and $y$ coordinates for the given MNPs, which, after linking together, yield the colored trajectory map. Apart from unintended global cumulative drift behavior, which was subtracted from all trajectories in the following analysis (cf. Figure~\ref{fig:drift_subtraction.pdf}), the trajectories show no visible directionality, as is expected from the random-walk character of Brownian motion.\\
Our goal was to examine the magnitude of the Brownian motion as a function of the MNP's DNA length.
It can be extracted from the tracking data by determining a particle's change in position $\Delta x$ and $\Delta y$ over lag-time $\tau$ (cf. Equation~\ref{eq:1}). Accordingly, when determining an MNP's MSD with increasing lag-time $\tau$, the slope of the data presents a means to quantify the MNP's diffusivity, i.e., the magnitude of its Brownian motion. In Figure~\ref{fig:2}~(c), exemplary drift-corrected MSD data obtained by tracking MNPs without DNA attachment (cf. Figure~\ref{fig:2}~(b)) are shown. Each black line corresponds to the MSD of a single tracked MNP. Some outliers can be observed; however, the general trend reveals an almost linear increase of MSD over time, with similar $D$ for individual curves. When interpreting the obtained distribution of MSD curves, the size distribution of observed MNPs has to be taken into account. This results from the dependency of $D$ on the MNP diameter $d$ (cf. Equation~\ref{eq:2}). \\
To account for anomalous subdiffusion in the observed motion of the MNPs, the MSD($\tau$) data were fitted with a power law of the form~\cite{Saxton1997, Hfling2013}:
\begin{equation}
\mathrm{MSD}(\tau)=A\cdot \tau^{\alpha},
\label{eq:3}
\end{equation}
with $A=4D$, including the diffusivity of an object moving only in two dimensions. The exponent $\alpha$ is equal to one for ideal Brownian motion~\cite{Saxton1997, Hfling2013}, but may differ depending on the experimental conditions. For the fitting procedure, we fixed $\alpha$ to range between 0.95 and 1.05 to avoid a result that is too far off from Brownian motion. Additionally, we removed outliers from the MSD($\tau$) data before fitting (details are provided in the Method section).\\
Fitting the MSD data for each tracked MNP in a single video, a distribution of $D$ was obtained from $D=A/4$. These distributions characterize the free 2D Brownian motion of MNPs functionalized with dsDNA of different lengths. To account for the uncertainties of the fitting parameters $D$ and $\alpha$, a Monte-Carlo approach has been taken to create histograms for $D$ and $\alpha$ for each MNP type (different DNA length). For each fit of the MSD($\tau$) data of a single MNP, a normal distribution of $D$ and $\alpha$ is considered. These distributions are added up after fitting the MSD($\tau$) data of all tracked MNPs within a recording, resulting in Monte-Carlo histograms for each parameter. The histograms for $\alpha$ are included in the Supporting Material. The MSD($\tau$) data, the corresponding histograms for $D$ and $\alpha$ are displayed for each investigated dsDNA length in Figure~\ref{fig:3}~(a)-(e), Figure~\ref{fig:3}~(f)-(j), and Figure~\ref{fig:3}~(k)-(o), respectively. To derive experimentally determined diffusion coefficient distributions, the Monte-Carlo histograms were fitted with lognormal distributions and summarized in Figure~\ref{fig:D_exp}.

Two trends are observable when comparing bare MNPs (0 bp DNA) with MNPs functionalized with maximum DNA strand length (\SI{2000}{\bp}): (1) The variance of the $f_{D,\mathrm{exp}}\left(D\right)$  is narrowing, and (2) the mean values of $f_{\mathrm{exp}}\left(D\right)$ are slightly but significantly decreasing. When looking at the intermediate DNA lengths (\SI{340}{\bp}, \SI{500}{\bp}, \SI{722}{\bp}), this trend persists, though the narrowing and shift of the $D$ distribution starts to become evident for \SI{500}{\bp} DNA strands attached to the MNPs. The $f_{\mathrm{exp}}\left(D\right)$ distributions for 0 bp and \SI{340}{\bp} DNA appear very similar; the same holds when comparing \SI{500}{\bp} with \SI{722}{\bp} DNA.
\\

This mirrors the trend predicted by the forward model in Figure~\ref{fig:D_model}. A closer quantitative comparison, however, reveals that the measured diffusion coefficients are systematically smaller than the values expected for single MNPs with a dsDNA corona. This already suggests that the tracked objects are not isolated MNPs but larger hydrodynamic units, i.e., MNP agglomerates.

\subsection{Inverse reconstruction of effective diameter distributions}\label{subsec: inv effect diameter}

\begin{figure}[h!]
\centering
\includegraphics{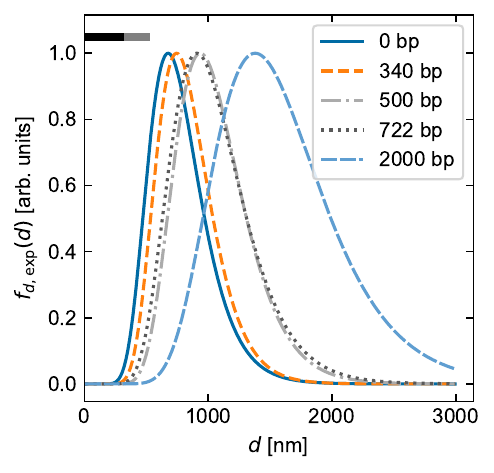}
\caption{Effective hydrodynamic diameter distributions $f_{d,\mathrm{exp}}\left(d\right)$ reconstructed from the experimental diffusion coefficient distributions $f_{D,\mathrm{exp}}\left(D\right)$ by inverse Stokes–Einstein transformation. The distributions shift monotonically to larger diameters with increasing dsDNA length and are systematically broader and larger than the single-MNP predictions in Figure~\ref{fig:D_model}, indicating MNP agglomeration. The black and gray bands indicate the Abbe limit.} 
\label{fig:d_exp}
\end{figure}

To quantify the hydrodynamic sizes actually probed in the tracking experiments, we use the fitted lognormal parameters $\mu_D$ and $\sigma_D$ and the transformation Equation~\ref{eq:f_d(d)} to reconstruct effective diameter distributions $f_{d,\mathrm{exp}}\left(d\right)$ for each MNP population. The reconstructed distributions are shown in Figure~\ref{fig:d_exp}. As in the model, the peaks of the distributions move to larger diameters as the dsDNA length increases.
At the same time, the reconstructed diameters are noticeably larger than the single-MNP predictions from Figure~\ref{fig:d_model}, and the distributions for the longest dsDNA, especially \SI{2000}{\bp}, develop pronounced high-diameter tails. Both features are consistent with dsDNA-mediated clusters rather than single MNPs.

\subsection{Consistency between hedgehog model and agglomeration}\label{subsec: hedgehog agglo}

\begin{figure}[h!]
\centering
\includegraphics{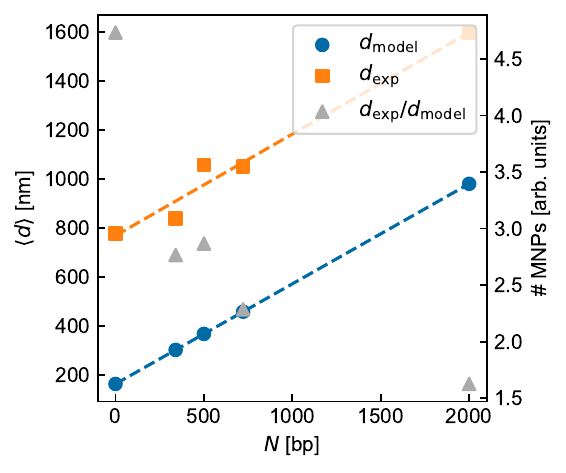}
\caption{Mean values of the diameter distributions from the attenuated hedgehog model (single MNPs, Figure~\ref{fig:d_model}) and from the reconstructed experimental distributions (Figure~\ref{fig:d_exp}) as a function of dsDNA length. Both sets of maxima follow the same monotonic trend with dsDNA length, while the offset quantifies the degree of MNP agglomeration.}
\label{fig:agglo}
\end{figure}

For a compact comparison, we extracted the mean values of the modeled effective diameter distributions and of the reconstructed experimental distributions and plotted them as a function of dsDNA length (Figure~\ref{fig:agglo}). The model mean values, corresponding to single MNPs decorated with an attenuated hedgehog corona, increase almost linearly with dsDNA length. The mean values of the reconstructed distributions follow the same monotonic trend but are shifted to larger values. When the experimentally obtained means are divided by the model-predicted means, the resulting dimensionless ratios provide a simple measure of the average agglomerate size. For bare MNPs, these ratios correspond to several MNPs per cluster; for \SI{2000}{\bp} dsDNA, they approach values much closer to unity. In other words, long dsDNA still promotes cluster formation, but the resulting assemblies are less compact and behave hydrodynamically more like a small number of MNPs tied together by flexible DNA tethers. The important point is that the attenuated hedgehog model is not contradicted by the observed agglomeration. Quite the opposite: it sets the underlying dependence of hydrodynamic size on dsDNA length, and the clusters simply shift this curve to larger apparent diameters. The agreement in trend between theory and experiment suggests that the dsDNA corona dominates the hydrodynamic scaling, even when several MNPs are involved.

\section{Discussion}\label{sec:discussion}
By analyzing the free 2D Brownian motion of MNPs using optical microscopy and an optimized particle tracking algorithm, we intended to identify the structural conformation of differently sized dsDNA strands attached to the MNP surfaces. Specifically, we wanted to answer the question of whether DNA strands are wrapping around the MNP (coiled state) or rather pointing straight away from the MNP surface (brushed state). According to our investigation of the MNPs' diffusivity and the associated hydrodynamic diameter, we are able to impose a strong argument for the case of a partially brushed DNA conformation. The mean hydrodynamic diameter is increasing by \SI{105}{\percent} for MNPs functionalized with \SI{2000}{\bp} DNA strands compared to non-functionalized MNPs, which is not explicable by DNA strands wrapping around the MNP. Therefore, we assume the effective hydrodynamic diameter using the attenuated hedgehog model, which is valid under the given experimental conditions.\\
The experimentally determined increase of the hydrodynamic diameter for \SI{2000}{\bp} dsDNA-MNPs correlates with the theoretical increase in the case of a partially brushed DNA conformation (adding the mean MNP radius and the mean end-to-end distance - of the attenuated hedgehog model - of the dsDNA strand). It also explains the observed narrowing of the diffusivity histograms when increasing the DNA length: Due to the comparably low MNP size, the impact of the DNA length becomes more pronounced when moving towards \SI{2000}{\bp} strands. Therefore, the effect of MNP size distribution, as examined by TEM measurements, diminishes when assuming that all DNA strands attached to the MNP surface stay the same in length.\\

The measurements presented here allow a direct comparison between the hydrodynamic response of DNA-functionalized MNPs and the behavior expected from a simple geometric model that treats the semiflexible dsDNA fragments as radially extended. Under the low-ionic-strength conditions used in our experiments, electrostatic repulsion keeps the DNA in a largely stretched configuration, so that the blunt ends of the fragments form an outer corona around each MNP. Since denaturing agents were absent, hybridization events by electrostatic interactions between the dsDNA molecules can be excluded as a cause for particle aggregation. In this regime, the effective hydrodynamic diameter grows approximately linearly with DNA contour length, a trend that is reproduced both in the analytical model and in the experimentally determined diffusion constants.\\

The comparison between measured and predicted size distributions provides a second important insight: although functionalization with dsDNA increases the apparent MNP size substantially, the degree of agglomeration remains modest. For all dsDNA lengths, the reversed diffusion-derived diameter distributions indicate only small clusters rather than large aggregates. This behavior is consistent with the expected interaction mechanism. In the absence of salt, the only attractive contribution arises from short-ranged blunt-end stacking between neighboring DNA brushes.~\cite{RomeroSanchez2022} Such interactions require both a favorable orientation and a sufficiently high local density of accessible ends. The moderate number of dsDNA strands per MNP in our system therefore limits the magnitude of the attraction and, in turn, the size of the resulting clusters.\\

The observed dependence on fragment length fits naturally into this picture. Longer DNA strands extend the steric reach of the brush and increase the likelihood of contact between neighboring coronas, which explains the systematic shift of the effective size distributions. At the same time, however, the blunt-end interaction remains very local; increasing the dsDNA length does not proportionally strengthen the attraction. As a consequence, the cluster sizes inferred from the diffusion measurements increase only mildly and eventually saturate. Even the longest dsDNA fragments examined here (\SI{2000}{\bp}) lead to hydrodynamic growth but only limited agglomeration.\\

It has to be noted that the presented investigation of diffusivity for DNA-functionalized MNPs is connected with some uncertainties resulting from the experimental conditions as well as the data evaluation. First of all, MNPs are free to move in 3D while being recorded. Thus, MNPs may move vertically with respect to the focal plane of the optical microscope, causing blurring or sharpening of the MNP image. This may deteriorate the accuracy of the applied MNP tracking algorithm, on which the determination of MSD is founded. Furthermore, MNPs might drift toward a certain direction due to unforeseen fluid currents or a tilt of the fluid cell. An attempt was made to subtract this drift in the MSD($\tau$) data; however, as can be seen, for instance, in Figure~\ref{fig:3}~(b), some single curves deviate from the typical behavior. These few outliers supposedly play only a minor role when determining the mean diffusivity of the MNPs. Nonetheless, when fitting the MSD($\tau$) data with Equation~\ref{eq:3}, we accounted for the uncertainties of the fitting parameters $D$ and $\alpha$, which are hard to quantify directly, by using a Monte-Carlo distribution approach. The resulting mean diffusivities and their deviations can, therefore, be considered to be statistically sound. This highlights the strength of our investigation compared to other single particle-based techniques: Tracking the Brownian motion of many MNPs simultaneously, ie., between 100 and 300 particles for each video,  allows for a large MSD($\tau$) data set, providing profound statistical significance for examining the MNPs' hydrodynamic radii and the connected conformation of grafted DNA strands.

Taken together, these findings show that the interplay between DNA stretching, blunt-end accessibility, and grafting density determines whether and to what extent DNA-functionalized MNPs form clusters under salt-free conditions. The agreement between the modeled effective diameters and the distributions recovered from the diffusion data demonstrates that the dominant contribution to the measured behavior is steric and hydrodynamic rather than strongly adhesive. At the same time, the small but systematic deviations from the single-particle predictions reveal the onset of weak, dsDNA-mediated aggregation.

\section{Conclusion}\label{sec:conclusion}
We have shown how TEM-based size statistics of bare MNPs, a physically motivated hedgehog model for dsDNA coronas under low-salt conditions, and optical tracking of Brownian motion can be combined into a single, coherent framework for describing dsDNA-functionalized MNPs from the single-particle to the cluster level.

The main conclusions are:
\begin{itemize}
    \item TEM reveals that the nominal \SI{300}{nm} streptavidin-coated MNPs in fact follow a lognormal size distribution with a mean diameter of about \SI{165}{nm} and significant polydispersity. This real core distribution must be taken into account when interpreting hydrodynamic measurements.
    \item Using an attenuated hedgehog model with $\gamma = 0.6$, the TEM-derived diameter distribution can be propagated to effective diameter distributions and diffusion coefficient distributions for dsDNA of \SIrange{340}{2000}{\bp}.
    \item Optical tracking yields diffusion coefficient distributions that decrease systematically with dsDNA length, in line with the model predictions. Inverse use of the Stokes-Einstein relation allows us to reconstruct effective diameter distributions from the measured diffusion data.
    \item The reconstructed effective diameters are larger than those expected for single MNPs, demonstrating that the tracked objects are dsDNA-mediated MNP agglomerates. Nevertheless, the maxima of the reconstructed and modeled distributions follow the same dependence on dsDNA length, which means that the hydrodynamic scaling set by the dsDNA corona is preserved.
\end{itemize}

Taken together, these observations provide a consistent picture in which dsDNA not only increases the hydrodynamic envelope of individual MNPs but also acts as a flexible connector that brings MNPs together into clusters. The analysis chain presented here, consisting of TEM, forward modeling, optical tracking, and inverse reconstruction, should be broadly applicable to other functionalized colloidal systems in which structure and hydrodynamics are intertwined, and the optical resolution limit masks the underlying core sizes.

\section{Experimental Section}\label{sec:methods}
\subsection{Synthesis of DNA strands and binding to magnetic particles}
Genomic DNA was isolated from Arabidopsis thaliana according to Edwards et al. (1991).\cite{edwards_simple_1991} DNA fragments of different lengths were produced by artificial polymerase chain reaction (PCR) using biotinylated forward primer (5‘ TCA ACC AAA TCC AAA TTC TAC ACA T 3‘) and Atto520 labelled reverse primer (\SI{340}{\bp}: 5’ ATA GGT ACG CGG AGC AAC AT 3’, \SI{500}{\bp}: 5’ TCG AAA GAA ATC GTT GAA GC 3’, \SI{722}{\bp}: 5’ CCA TTC TGT TAT GGC GGA AGA ACA C 3’, \SI{2000}{\bp}: 5’ TCT ACT TCC ACA TCC ATT CTT TC 3’) with genomic DNA as template. The PCR reaction products were separated by agarose gel electrophoresis, and the PCR product was isolated 
from the gel to remove excess primer oligonucleotides (Nucleo Spin\textsuperscript{\tiny\textregistered} Gel and PCR Clean Up System, Machery Nagel). The DNA concentration was determined with the NanoPhotometer\textsuperscript{\tiny\textregistered} N60/N50 (Implen).\\

Biotinylated DNA was coupled to streptavidin-coated \SI{300}{nm}  magnetic particles from Ademtech Pessac, France according to manufacturer’s instructions. Briefly, \SI{10}{\micro\l} magnetic particles (\SI{5}{\milli\g\per\milli\l}) were washed twice with immobilization buffer (pH 7). \SI{2}{\micro\g} modified DNA suspended in a total volume of \SI{10}{\micro\l} was added to the magnetic particles and gently shaken in an overhead rotator for 30 min at room temperature. Particles were washed three times with immobilisation buffer and finally resuspended in H$_2$O to a final concentration of \SI{1}{\nano\g} particles per \si{\micro\l} solution.\\
\subsection{Experimental setup}
Suspensions of DNA-functionalized particles (and bare particles for control) were pipetted into wells of a microtiter plate. These wells were placed below an optical brightfield microscope. Particles that moved freely, i.e., far away from the well's walls, were placed in the focal plane of the microscope. Videos of \SI{120}{\second} duration were recorded for each particle type (different DNA length), with a frame rate of 60 frames per second and at an image resolution of 2048 pixels $\times$ 1536 pixels using a high-resolution camera (DMK 33UX265 with Sony IMX265 sensor).\\

\subsection{Particle tracking}

        60 fps bright-field microscopy videos were analyzed using modified functions of the Python library trackpy 0.7. In a first step, the $xy$ position  $\mathbf{r}  \left(t\right)=\left( x  \left(t\right), y  \left(t\right)\right)$ of all particles for each relevant frame (at time $t$) was determined. The quality of the locator parameters was improved by taking the subpixel accuracy distribution as well as the mass (inverted brightness) distribution into account. Optimal particle detection parameters were found to be \verb|size = 11 px|, \verb|minmass = 300|, \verb|threshold = 1|, \verb|separation = 23 px|. The separation value, which is more than two times higher than the average particle size, prevents agglomerates from being counted as multiple particles. The uncertainty of particle location was estimated as $\sigma_{\mathrm{pos}}=1~\mathrm{px}$ for the $x$ and $y$ coordinates, respectively.\\

        Particles of consecutive frames are linked to a respective trajectory. Thereby, it was considered that the particles may leave and come back to the focal plane. To prevent the particle from being counted twice, a retrieval algorithm was used whereby if a particle was lost, it was searched for \verb|memory = 20 frames| in a radius of \verb|searchrange= 20 px| from its last position. If a particle is located again, the two trajectories are merged and considered as one particle. As the separation of located particles is more than two times their size, we can be sure that the retrieved particle is the original particle, and it is therefore valid to merge the trajectories. \\

        Further, we consider only particles with a frame length of at least 1800 frames (\SI{30}{s}), removing spurious trajectories which do not contribute significantly to the MSD. Additionally, particles that move close to the border of the field of view, which could be counted multiple times by reentering the field of view, are removed from the data.\\

        To obtain the random motion of the particles, each trajectory $\mathbf{r}\left(t\right) = \left(x\left( t\right), y\left( t \right) \right)$ needs to be corrected from a global drift $\mathbf{d}\left(t\right) = \left(d_x \left( t\right), d_y \left( t \right) \right)$, resulting to a drift corrected position $\mathbf{r}'\left(t\right) = \left( x'\left( t\right), y'\left( t\right)\right)$ of the particles in each frame.
        \begin{align}
            \mathbf{r}'\left(t \right) &= \mathbf{r}\left(t\right) - \mathbf{d}\left(t \right).
        \end{align}
        Whereby the global drift $\mathbf{d}$ is calculated by the frame-to-frame distance $\Delta\mathbf{r} \left(t\right) = \left(\Delta x \left( t \right), \Delta y \left( t \right) \right)$ of the particles' positions for each time $t$ during the duration of the video.
                \begin{align}
            \Delta \mathbf{r} \left(t\right) &= \mathbf{r} \left(t\right) - \mathbf{r} \left(t -1\right).
\end{align}
The frame-to-frame distance is averaged over all particles for each time frame, respectively, whereby $N\left(t\right)$ denotes the number of particles at the respective time (frame), leading to $\Delta \mathbf{\bar{r}}\left(t\right) = \left(\bar{\Delta x}\left( t \right)  , \bar{\Delta y}\left( t \right)\right)$.
\begin{align}
                \Delta \mathbf{\bar{r}} \left(t\right) &=\frac{1}{N\left(t\right)}\sum \Delta \mathbf{r} \left(t\right).
\end{align}
   Next, the cumulative sum of $\bar{\Delta \mathbf{r}}\left(t\right)$ results to the global time dependent drift $\mathbf{d}\left( t \right) = \left(d_x\left( t\right), d_y\left( t\right)\right)$. Note that the variable $k$, which is interchangeable with $t$, was introduced to describe the cumulative sum.
\begin{align}
           \mathbf{d}\left( t \right) &=\sum_{k=1}^t \bar{\Delta \mathbf{r}} \left(k\right).
        \end{align}
        According to the calculations above the corresponding uncertainties for the drift-corrected trajectories $\mathbf{r}' \left(t\right)$ are obtained as follows.

    \begin{align}
        \sigma_{\Delta x} \left(  t \right) &= \sigma_{\Delta y} \left( t \right) =\sqrt{\sigma^2_{\mathrm{pos}}+\sigma^2_{\mathrm{pos}}} = \sqrt{2} \sigma_{\mathrm{pos}},\\
        \sigma_{\bar{\Delta x}}\left( t \right) &= \sigma_{\bar{\Delta y}}\left( t \right) = \frac{\sigma_{\Delta x} \left( t \right)}{\sqrt{N\left( t \right)}} = \frac{\sqrt{2}\sigma_{\mathrm{pos}}}{\sqrt{N\left(t\right)}},\\
        \sigma_{d_x} \left( t \right) &= \sigma_{d_y} \left( t \right) =\sqrt{\sum_{k=1}^t \sigma_{\bar{\Delta x}}\left( k \right)^2} = \sqrt{\sum_{k=1}^t \frac{2\sigma^2_{\mathrm{pos}}}{N\left(k\right)}},\\ 
        \sigma_{x'} \left( t\right) &= \sigma_{y'} \left( t\right) = \sqrt{\sigma_{\mathrm{pos}}^2+ \sigma_{d_x} ^2\left( t \right) } = \sqrt{\sigma_{\mathrm{pos}}^2+ \sum_{k=1}^t \frac{2\sigma^2_{\mathrm{pos}}}{N\left(k\right)}}.
    \end{align}\\

        The mean squared displacement (MSD) is determined by first taking the difference of the particles' drift corrected position $\mathbf{r}'_i = \left(x_i, y_i\right)$ separated by lag time $\tau$, whereby $i \in \mathcal{I}\left( \tau \right)$ with $\mathcal{I}(\tau) = \left\{ i \;\middle|\; \mathbf{r}_i \text{ and } \mathbf{r}_{i+\tau} \text{ exist} \right\}$.
        \begin{align}
\Delta \mathbf{r}_i \left( \tau \right) &= \mathbf{r}_{i+\tau}- \mathbf{r}_i = \left(\Delta x_i \left( \tau \right), \Delta y_i \left( \tau \right) \right).
\end{align}
        At this point of the analysis, pixel values are converted to real position values in µm using the experimentally determined px-to-µm conversion factor \SI[round-mode=places,round-precision=4]{0.1733712121}{\micro\meter\per pixel}. To get the squared displacement $\langle \Delta \mathbf{r}^2\rangle \left( \tau \right)$, the positional difference vector $\Delta r_i \left( \tau \right)$ are squared and summed componentwise for each $i$ and divided by $M\left( \tau \right) = \left| \mathcal{I}\left( \tau \right) \right|$, which represents the number of possible pairs of two positions of a particle separated by $\tau$,         
 \begin{align}       
 \langle \Delta \mathbf{r}^2\rangle \left( \tau\right) &= \frac{1}{M\left( \tau\right)}\sum_{i\in\mathcal{I\left( \tau \right), j}} \left( \Delta \mathbf{r}_{i,j}\right)^2 \left( \tau \right)= \left(\langle\Delta x^2\rangle\left(\tau\right), \langle\Delta y^2\rangle\left(\tau\right)\right)~\mathrm{with}~j\in\left\{x,y\right\}.
 \end{align}

$\mathrm{MSD}\left(\tau\right)$ is obtained by summing up the components of $\langle \Delta \mathbf{r}^2  \rangle\left(\tau\right)$,
\begin{align}
\mathrm{MSD}\left( \tau \right) &= \langle \Delta x^2\rangle \left( \tau\right) + \langle \Delta y^2\rangle \left( \tau\right).
        \end{align}
        Corresponding uncertainties are calculated as follows.
        \begin{align}
        \sigma_{\Delta \mathbf{r}_{i,j}}(\tau) &= \sqrt{\sigma^2_{\mathbf{r},i+\tau,j}+\sigma^2_{\mathbf{r},i,j}},\\
\sigma_{ \Delta \mathbf{r}_{i,j} ^2}\left( \tau \right) &\approx 2\left|\Delta \mathbf{r}_{i,j} \left( \tau \right) \right| \sigma_{\Delta \mathbf{r}_{i,j}}\left( \tau \right),\\
\sigma_{\langle \Delta \mathbf{r}_j^2 \rangle } \left( \tau \right) &= \frac{1}{M\left( \tau \right)} \sum_{i \in \mathcal{I}\left( \tau \right)}\sigma_{\Delta \mathbf{r}^2_{i,j}}\left( \tau \right),\\
\sigma_{\mathrm{MSD}}\left(\tau\right) &= \frac{\sqrt{\left( \sigma_{\langle \Delta x^2\rangle } \left( \tau \right) \right)^2+\left( \sigma_{\langle \Delta y^2\rangle } \left( \tau \right) \right)^2}}{\sqrt{T_{\mathrm{eff}} \left( \tau \right)}},\\
T_{\mathrm{eff}}\left(\tau\right) &= \begin{cases}
                \displaystyle \frac{6 (T-\tau)^2 \tau}{2T - \tau + 4 T \tau^2 - 5 \tau^3}, & \tau \le T/2,\\[2ex]
                \displaystyle \frac{1}{1 + \frac{(T-\tau)^3 + 5\tau - 4 (T-\tau)^2 \tau - T}{6 (T-\tau)\tau^2}}, & \tau > T/2.
            \end{cases}
\end{align}
Hereby, $T_{\mathrm{eff}}$ describes the effective number of statistically independent measurements for a trajectory of length $T$~\cite{Qian1991SinglePT}.\\

        As also stuck particles or dirt were located by trackpy we need to filter out trajectories that show no significant MSD. The threshold value was determined empirically. Trajectories that have an increase of $\log_{10} \mathrm{MSD}$ smaller than $O_{\mathrm{min}} = 0.045$ over $\Delta t = t_1 -t_0=\SI{1.65}{ s}$ were neglected.
        \begin{align}
            \Delta \mathrm{MSD} &= \mathrm{MSD}\left(t_1\right)-\mathrm{MSD}\left(t_0\right),\\
            \log_{10}\left(\Delta \mathrm{MSD}\right) &< O_\mathrm{min}. \label{eq:filter}
        \end{align}   

        To gain information of the diffusivity $D$ of the particles, the logarithmic form of the  power law $\log_{10}\left( \mathrm{MSD}\left( \tau \right)\right) = \log_{10}\left(A\right) +\alpha \log_{10}\left(\tau\right)$, with $A = 2\cdot d \cdot D,~d=2$ was fitted to the $\mathrm{MSD}\left(\tau\right)$ data. Hereby, the lag time dependent uncertainty  $\sigma_{\mathrm{MSD}}\left(\tau\right)$ was used as a weighting factor for the fit, as described below, resulting in the residual $S$ which is minimized in the fitting procedure, whereby the Trust Region Reflective algorithm was used.
        \begin{align}
           \log \mathrm{MSD}\left(\tau\right) &= \log A + \alpha \log \tau, \label{eq:log_power}\\
            \sigma^2_{\log \mathrm{MSD}}\left(\tau\right) &=\frac{\sigma_{\mathrm{MSD}}\left(\tau\right)}{\mathrm{MSD}\left(\tau\right)},\\
            w\left(\tau\right) &= \frac{1}{\sigma^2_{\log \mathrm{MSD}}\left(\tau\right)},\\
             S &= \sum^{\tau_{\mathrm{max}}}_{n=0}\left[\left(\log_{10}\left(\mathrm{MSD}\right)\left(\tau_n\right)-\log_{10}\left(\hat{A}\right) +\hat{\alpha} \log_{10}\left(\tau_n\right)\right)\cdot w\left(\tau_n\right)\right]^2,\\
            \sigma_{\hat{\theta}} &=\sqrt{C_{ii}}, \quad \hat{\theta} = \left(\hat{A}, \hat{\alpha}\right).
        \end{align}
        Respective uncertainties of the fitting parameter are obtained from the diagonal covariance matrix $C$.\\

    In the next step, Monte-Carlo (MC)-histograms of $D=\hat{A}/4$ depending on the dsDNA length were created. By doing this we consider the uncertainty of the respective fitting parameters in the following determination of the hydrodynamic radius reasonably. The MC-histograms were calculated with a sample size of 10000 and a $3\sigma$-interval.

\section*{Acknowledgments} 
CJ, YS, RH, and AE acknowledge project and scientific infrastructure funding by the German Research Foundation (DFG) under the project numbers 361396165, 514858524.

\section*{Conflict of Interest}
The authors declare no conflict of interest.

\section*{Author Contributions}
CJ: data curation, formal analysis, investigation,  methodology, software, validation, visualization, and writing - original draft.\\
FS-M: investigation, methodology.\\
YS: software, validation.\\
MW: preparation, investigation, methodology.\\
K-JD: conceptualization, funding acquisition, project administration, resources, supervision.\\
IE: microstructural analysis, methodology.\\
RH: project administration, supervision, visualization, writing - original draft.\\
AE: funding acquisition, project administration, resources, supervision.\\
AH: conceptualization, funding acquisition, project administration, resources, supervision, writing – original draft.\\
All authors contributed to writing - review \& editing.

\bibliographystyle{unsrt}
\bibliography{bibliography}

\newpage
\section*{Supporting Information} 
\renewcommand{\thefigure}{S\arabic{figure}}
\renewcommand{\thetable}{S\arabic{table}}
\setcounter{figure}{0}
\setcounter{table}{0}

\begin{figure}[h!]
\centering
\includegraphics{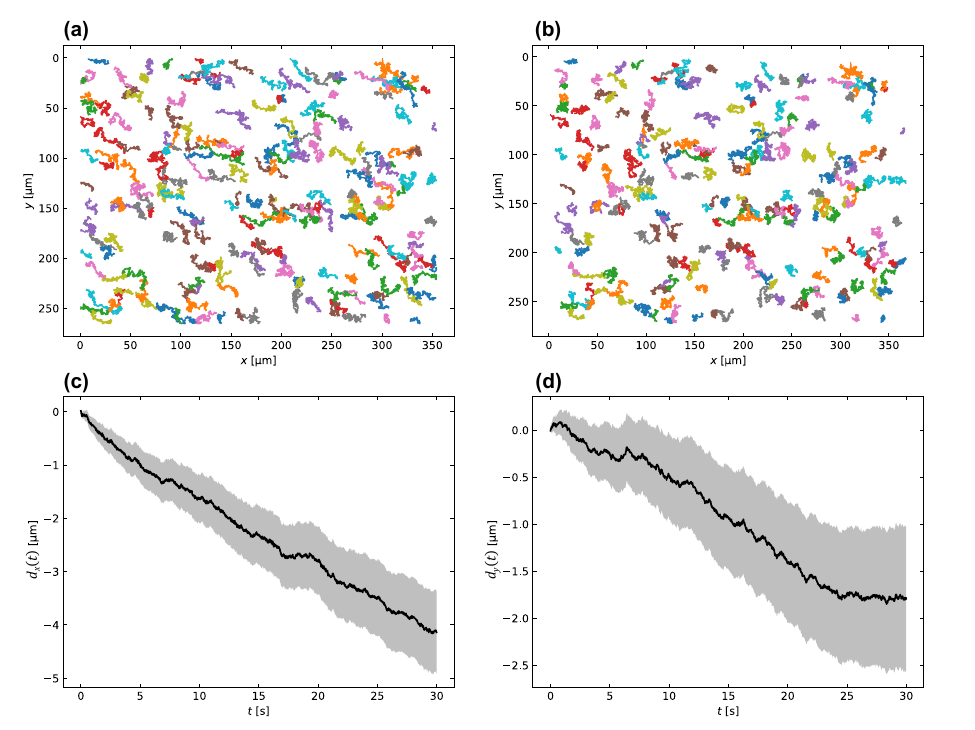}
\caption{Example for  (a) raw trajectories $\mathbf{r}\left(t\right)$, global cumulative drift (c) $d_{x}\left(t\right)$ and (d) $d_{y}\left(t\right)$, (d) drift corrected trajectories $\mathbf{r}'\left(t\right)$.} 
\label{fig:drift_subtraction.pdf}
\end{figure}

Exemplary plots in the supporting information are given for dsDNA-coupled MNPs with a dsDNA length of \SI{722}{\bp} if not stated otherwise.\\ 
To minimize the effects of a lateral global drift $\mathbf{d}\left(t\right)$ on the MSD, the drift was subtracted from the trajectory data, yielding drift-corrected trajectory data. Raw trajectory data is shown in Figure~\ref{fig:drift_subtraction.pdf}~(a). The cumulative global drift $d_{x}\left(t\right)$ (Figure~\ref{fig:drift_subtraction.pdf}~(c)) and $d_{y}\left(t\right)$ (Figure~\ref{fig:drift_subtraction.pdf}~(d)) are subtracted from the raw trajectory data to obtain drift-corrected trajectories (Figure~\ref{fig:drift_subtraction.pdf}~(b)). The drift uncertainties are given by the gray area.

\begin{figure}[h!]
\centering
\includegraphics{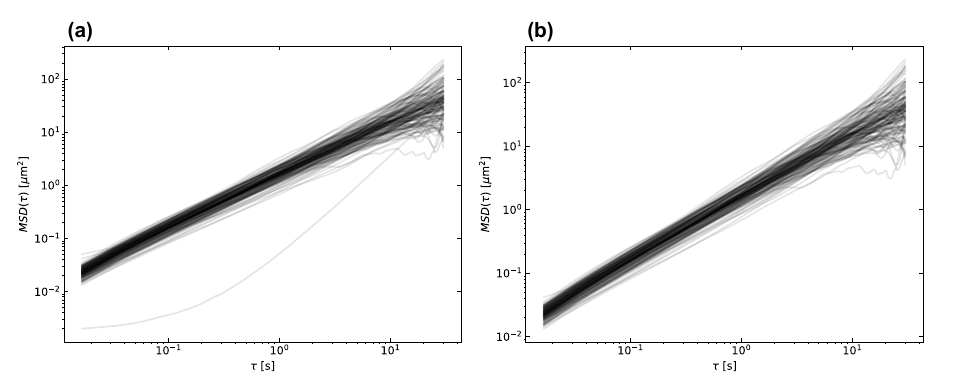}
\caption{Example for $\mathrm{MSD}\left(\tau\right)$ data (a) before  and (b) after  removal of measurement artifacts.} 
\label{fig:artefact_removal.pdf}
\end{figure}

Figure~\ref{fig:artefact_removal.pdf} illustrates the MSD data filtering based on Equation~\ref{eq:filter}. The MSD data of each particle trajectory was fitted using the logarithmic power law (cf. Equation~\ref{eq:log_power}.

\begin{figure}[h!]
\centering
\includegraphics[scale=0.9]{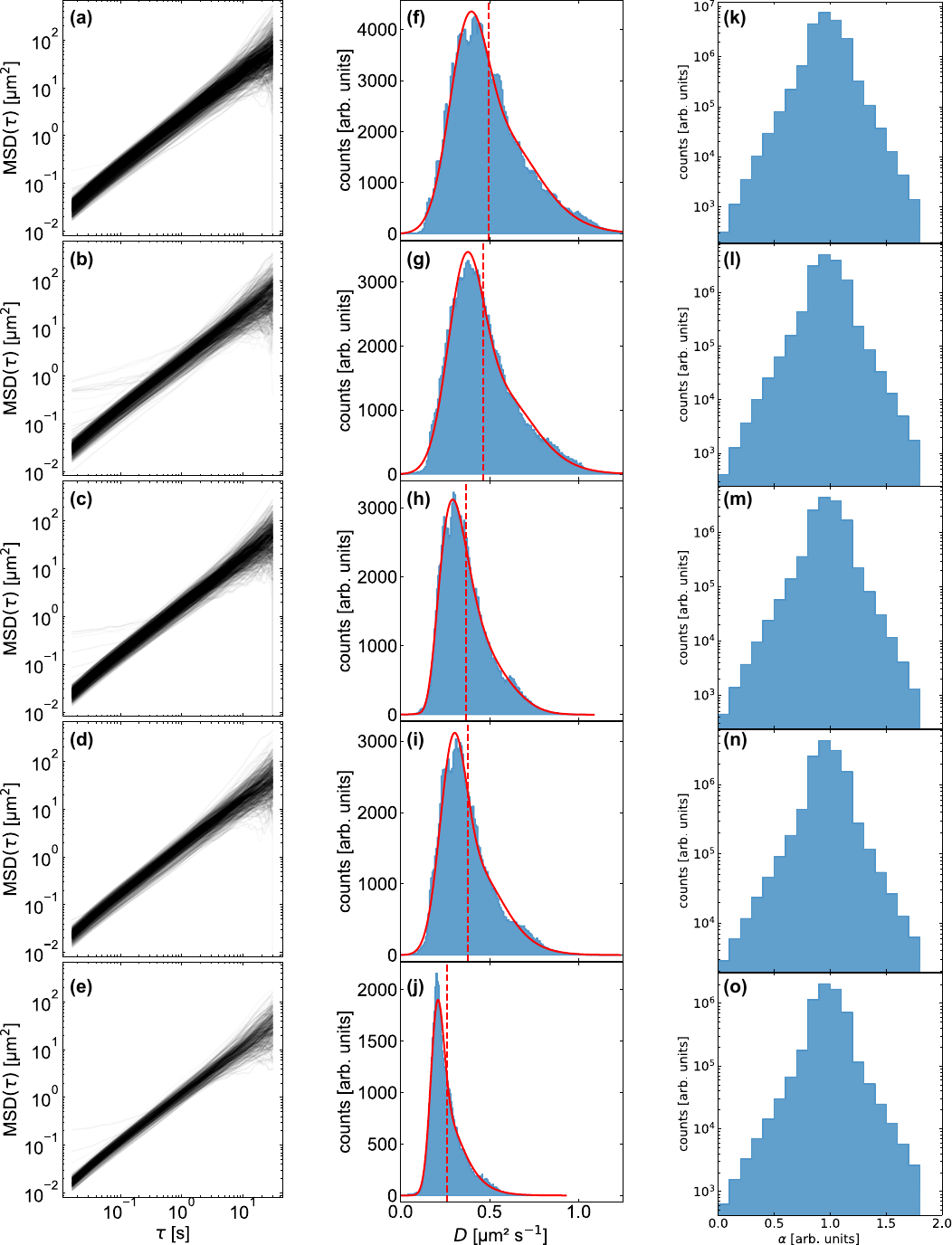}
\caption{Analysis of diffusivity $D$ for spherical MNPs with a nominal diameter of \SI{300}{\nano\meter} in dependence on the length of attached dsDNA strands (0 bp to \SI{2000}{\bp}). (a) - (e) Measured mean squared displacement (MSD) of MNPs as a function of lag time $\tau$ for different DNA lengths. (f) - (j) Histograms of $D$ obtained from fitting the MSD data for each MNP of specified DNA length. The mean of $D$ is visibly shifting toward smaller values as the DNA length increases, indicating a significantly increased hydrodynamic radius of MNPs due to a brushed arrangement of DNA strands within the attenuated hedgehog model. (k)-(o) corresponding Monte-Carlo histogram of the $\alpha$ values of the fit.} 
\label{fig:3}
\end{figure}

An overview of the MSD plots (black) for the respective dsDNA lengths is shown in Figure~\ref{fig:3}~(a)-(e). The diffusivity $D$, derived from the fitting parameter $A$, is shown respectively as a Monte-Carlo histogram in Figure~\ref{fig:3}~(f)-(j)  (blue). The Monte-Carlo histograms were fitted with a lognormal function (red) within the attenuated hedgehog model to determine the mean value of diffusivity, which is depicted by a dashed line.  Further, the Monte-Carlo histograms of the fitting parameter $\alpha$ are shown for each dsDNA length in Figure~\ref{fig:3}~(k)-(o). From top to bottom, the dsDNA length is increasing, i.e., (a), (f), (k) are referring to a dsDNA length of \SI{0}{\bp} and (e), (j), (o) are referring to \SI{2000}{\bp}. Visualizing the fitting parameter $\alpha$ as a Monte-Carlo histogram leads to counts outside of the fit parameter boundaries. Whereby these outlier counts are orders of magnitude less than the maximum around $\alpha=1$.

\end{document}